\documentclass[10pt, twocolumn]{article}

\usepackage{article_macros}
\usepackage[top = 1in, bottom = 1in, left = 0.7in, right = 0.7in]{geometry}
\usepackage{algorithm}
\usepackage{algpseudocode}
\usepackage[tt = false]{libertinus}

\newcommand{\cnot}{\textsc{cnot}}

\algrenewcommand{\Require}{\item[\textbf{Input:}]}
\algrenewcommand{\Ensure}{\item[\textbf{Output:}]}
\algrenewcommand{\alglinenumber}[1]{\footnotesize #1.}

\title{Edge-Local and Qubit-Efficient Quantum Graph Learning \\ for the NISQ Era}
\author{
    Armin Ahmadkhaniha \\ {\tt ahmadkha@mcmaster.ca} \and
    Jake Doliskani \\ {\tt jake.doliskani@mcmaster.ca} \and
    Department of Computing and Software, McMaster University
}

\date{}

\begin{document}
\maketitle

\begin{abstract}
    Graph neural networks (GNNs) are a powerful framework for learning representations from graph-structured data, but their direct implementation on near-term quantum hardware remains challenging due to circuit depth, multi-qubit interactions, and qubit scalability constraints. In this work, we introduce a hybrid quantum graph learning architecture designed explicitly for unsupervised learning in the noisy intermediate-scale quantum (NISQ) regime.

    Our approach combines a variational quantum feature extraction layer with an edge-local and qubit-efficient quantum message-passing mechanism inspired by the Quantum Alternating Operator Ansatz (QAOA) framework. The message-passing operation is decomposed into pairwise interactions along graph edges using standard single- and two-qubit gates. For a graph with $N$ nodes and $n$-qubit feature registers, this reduces the number of qubits required at one time from $Nn$ to at most $2n$. We train the model using the Deep Graph Infomax objective to perform unsupervised node representation learning. The external class labels are not used during graph construction or training and are used only to evaluate the learned embeddings. Experiments on the Cora citation network and the Phase 3 release of the 1000 Genomes Project show that the quantum edge interaction contributes to the quality of the learned node representations.
\end{abstract}

\vspace*{4mm}

\section{Introduction}

Graph-structured data arise naturally across many scientific domains, including social networks, molecular chemistry, genomics, and citation analysis. In recent years, machine learning methods---particularly graph neural networks (GNNs)---have become a standard tool for extracting meaningful representations from such data, enabling tasks such as classification, clustering, and unsupervised representation learning \cite{kipf2016semi, gao2018large, schlichtkrull2018modeling, gilmer2017neural, xu2018powerful, scarselli2008graph}.

Despite this success, learning effective representations for large, high-dimensional graphs remains computationally demanding. Classical graph neural networks rely on repeated aggregation and transformation of node features, which can become costly as graph size and feature dimensionality grow. These challenges have motivated interest in alternative computational paradigms that may offer new ways to represent and process complex, structured data.

Quantum machine learning (QML) investigates the use of parameterized quantum circuits as trainable models for classical or quantum data \cite{biamonte2017quantum, schuld2021machine, farhi2018classification, havlivcek2019supervised}. However, current quantum hardware operates in the noisy intermediate-scale quantum (NISQ) regime, characterized by limited qubit counts, short coherence times, and restricted circuit depth. As a result, practical QML algorithms for near-term hardware should limit their qubit requirements and circuit depth and use operations that can be efficiently implemented on available quantum devices \cite{cerezo2021variational, preskill2018quantum, bharti2022noisy}.

These constraints are particularly challenging for graph learning. Classical graph convolutional networks (GCNs) rely on message passing driven by the adjacency structure of the graph, repeatedly aggregating information from neighboring nodes through linear transformations and nonlinearities \cite{kipf2016semi, gilmer2017neural, hamilton2017inductive}. Translating this paradigm directly to the quantum setting is challenging because encoding graph structure into quantum circuits can require interactions between many qubits, conditional operations, or high-depth constructions that scale with graph size. Such requirements can quickly exceed the capabilities of NISQ-era devices.

Several quantum graph learning models have been proposed in recent years. Early work introduced quantum graph neural networks in which graph structure is encoded through Hamiltonian-based variational evolutions \cite{verdon2019quantum}. Subsequent approaches developed quantum graph convolutional architectures inspired by classical GCNs, demonstrating promising performance on graph-level classification tasks \cite{zheng2021quantum, zheng2024quantum, ye2025quantum}. Other architectural variants have explored symmetry-aware and equivariant graph constructions \cite{mernyei2022equivariant}.

Some proposed quantum graph-learning constructions rely on global operations over many nodes or circuit components such as multi-controlled unitaries, which can be expensive to implement on near-term devices \cite{wang2023quantumgcn, chen2025jet}. In particular, an $n$-controlled unitary gate requires $O(n^2)$ two-qubit gates without ancilla qubits, and still $O(n)$ two-qubit gates even with $O(n)$ ancilla, making such constructions sensitive to noise and compilation overhead \cite{barenco1995elementary}.

To reduce these hardware requirements, a range of hybrid quantum--classical graph learning approaches have also been explored. Such models integrate quantum circuits as components within broader classical pipelines, often using classical processing before or after the quantum subroutine \cite{ai2024quantum, chen2021hybrid, tuysuz2021hybrid, salamatov2025quantum}. These hybrid decompositions reduce quantum resource requirements and make near-term implementation more practical.

In many hybrid architectures, quantum circuits are used for local feature extraction while classical algorithms perform neighborhood aggregation. In this case, the quantum circuit processes the node features, but the relation between neighboring nodes is still handled by a classical message-passing operation. This motivates us to consider whether the graph interaction itself can be implemented through a small quantum circuit without requiring a quantum register for every node in the graph.

\paragraph{Our contribution.}
In this work, we present a hybrid quantum graph learning framework designed explicitly for unsupervised learning on NISQ-era hardware. Our approach replaces global message-passing operations with a qubit-efficient, edge-local quantum message-passing mechanism inspired by the Quantum Alternating Operator Ansatz (QAOA) \cite{farhi2014quantum, hadfield2019quantum}. The quantum encoder consists of two main components: a variational quantum feature extraction circuit applied independently to each node, and a quantum message-passing layer that encodes graph structure through local interactions between pairs of neighboring nodes. The resulting embeddings are trained using the Deep Graph Infomax objective.

Our design avoids multi-controlled unitaries and uses standard single-qubit rotations and two-qubit entangling operations. Instead of representing all nodes on the quantum device at the same time, message passing is applied edge-by-edge. If each node is represented by an $n$-qubit feature register, the feature extraction step requires $n$ qubits and each edge-local message-passing step requires at most $2n$ qubits. Therefore, the number of qubits required at one time is independent of the number of nodes in the graph.

We evaluate our model on two datasets: the Cora citation network and the Phase 3 release of the 1000 Genomes Project. We use the Deep Graph Infomax objective for unsupervised representation learning \cite{velickovic2019deep} and compare the learned embeddings with classical, graph-only, hybrid, and interaction-removed baselines. The external labels are not used during training. They are used only after training to evaluate how well the learned representations agree with the known classes. 

The source code for our implementation is publicly available at \url{https://github.com/ArminAhmadkhaniha/QGCNlib}.

\section{Preliminaries}
\label{sec:prelims}

\subsection{Parameterized Rotations}

A general single-qubit unitary, up to a global phase, can be expressed using a set of rotation angles. The standard single-qubit rotation gates used in this paper are
\begin{align*}
    R_X(\theta) &= e^{-i \theta X / 2} = \cos(\theta/2) I - i \sin(\theta/2) X, \\
    R_Y(\theta) &= e^{-i \theta Y / 2} = \cos(\theta/2) I - i \sin(\theta/2) Y, \\
    R_Z(\theta) &= e^{-i \theta Z / 2} = \cos(\theta/2) I - i \sin(\theta/2) Z,
\end{align*}
where $X$, $Y$, and $Z$ are the Pauli operators. A general single-qubit rotation with learnable parameters $(\alpha, \beta, \gamma)$ is defined as
\begin{equation*}
    R(\alpha, \beta, \gamma) = R_Z(\alpha) R_Y(\beta) R_Z(\gamma).
\end{equation*}
In matrix form, $R(\alpha, \beta, \gamma)$ can be expressed as
\[
    \renewcommand{\arraystretch}{1.5}
    R(\alpha, \beta, \gamma) =
    \begin{bmatrix}
        e^{\frac{-i(\alpha+\gamma)}{2}} \cos\big(\frac{\beta}{2}\big) & -e^{\frac{-i(\alpha - \gamma)}{2}} \sin\big(\frac{\beta}{2}\big) \\
        e^{\frac{i(\alpha - \gamma)}{2}} \sin\big(\frac{\beta}{2}\big) & e^{\frac{i(\alpha + \gamma)}{2}} \cos\big(\frac{\beta}{2}\big)
    \end{bmatrix}.
\]

In quantum machine learning, a quantum circuit prepares a state $\ket{\psi_\theta}$ using gates parameterized by a set of trainable parameters $\theta$. Classical information is then obtained by measuring observables of the prepared state. In this work, we use expectation values of Pauli-$Z$ operators. For an $n$-qubit state, the expectation value of $Z$ on qubit $j$ is
\begin{equation*}
    \lrang{Z_j} = \bra{\psi_\theta} Z_j \ket{\psi_\theta},
\end{equation*}
where $Z_j$ denotes the Pauli-$Z$ operator acting on qubit $j$ and the identity on the remaining qubits.

A variational quantum circuit (VQC) consists of an input encoding, a parameterized unitary $U(\theta)$, and a measurement step. The measured values are used to construct a classical loss function, and the parameters $\theta$ are updated using a classical optimization method to minimize this loss \cite{schuld2018supervised}. We use this variational approach in our model.

\subsection{Graph Convolutional Neural Networks (GCNs)}

The graph convolutional network (GCN) model used in this work is based on the formulation introduced by Kipf and Welling~\cite{kipf2016semi}. Given a graph $G = (V, E)$ with $N$ nodes, feature matrix $X \in \mathbb{R}^{N \times d}$, and adjacency matrix $A \in \{0,1\}^{N \times N}$, the forward propagation rule for a single GCN layer is defined as

\begin{equation}
    \label{eq:classical-gcn}
    H^{(l+1)} = \sigma\big( D^{-1/2} \widetilde{A} D^{-1/2} H^{(l)} W^{(l)} \big),
\end{equation}

where $\widetilde{A} = A + I$ is the adjacency matrix with self-loops, $D$ is the corresponding degree matrix, and $H^{(0)} = X$. If $H^{(l)} \in \mathbb{R}^{N \times d_l}$, then $W^{(l)} \in \mathbb{R}^{d_l \times d_{l+1}}$ is the trainable weight matrix at layer $l$, and $\sigma$ is a nonlinear activation function.

The term $H^{(l)} W^{(l)}$ performs a learnable feature transformation, while the normalized adjacency matrix $D^{-1/2} \widetilde{A} D^{-1/2}$ propagates information between neighboring nodes. Each GCN layer therefore combines feature transformation with neighborhood aggregation.

One or more GCN layers can be used to produce a node embedding matrix. These embeddings can then be used for downstream tasks such as classification and clustering, or as the node representations optimized by an unsupervised objective such as Deep Graph Infomax.

\section{A Hybrid Quantum Graph Learning Algorithm}
\label{sec:methods}

Let $G=(V,E)$ be the input graph with $N$ nodes, adjacency matrix $A\in \{0,1\}^{N\times N}$, and node feature matrix $X \in \R^{N\times d}$, where each row represents the features of a single node. Our proposed quantum graph learning algorithm consists of three main components:
\begin{enumerate}[itemsep = 0mm]
    \item Quantum feature extraction via a variational quantum circuit,
    \item A single-layer quantum message passing unitary acting as a graph-structured message-passing operation,
    \item An unsupervised Deep Graph Infomax (DGI) objective applied to the resulting node embeddings.
\end{enumerate}
We describe each of these components in detail in the following sections.

\subsection{Quantum Feature Extraction}
\label{sec:qfe}

Let $\bm{x} = (x_0, x_1, \dots, x_{d - 1})$ be any row of the feature matrix $X$. Then $\bm{x}$ can be expressed as a quantum state using amplitude encoding~\cite{larose2020robust} as follows. Let $n = \ceil{\log_2 d}$. Using amplitude encoding on $n$ qubits, we prepare the state
\begin{equation*}
    \ket{\psi_{\bm{x}}} = \frac{1}{\opnorm{\bm{x}}}\sum_{j=0}^{d-1} x_j \ket{j}.
\end{equation*}

To extract features, we apply a variational circuit consisting of $L$ layers of an entangling block, for some $L \geq 1$. Such circuits alternate single-qubit rotations with entangling gates. For example, following Schuld \textit{et al.}~\cite{schuld2020circuit}, an entangling block consists of a layer of single-qubit rotation gates applied to each qubit, followed by a layer of controlled two-qubit gates connecting qubit pairs at a fixed range. 
% Denoting by $C_{c}(U_{t})$ a controlled-$U$ gate with control qubit $c$ and target qubit $t$, such a block can be written as
% \begin{equation}
%     \label{eq:vqc}
%     B = \prod_{k=0}^{\left\lfloor n/r \right\rfloor -1} C_{c_k} (U_{t_k}) \prod_{j=0}^{n-1} U_j.
% \end{equation}
% Here, the qubit pairs $(c_k,t_k)$ depend on the range parameter $r$, which can take values $r = 1, 2, \dots, n-1$.

In practice, we use general $R(\alpha_j,\beta_j,\gamma_j)$ gates on each qubit, followed by \cnot{} gates entangling qubit $k$ with qubit $(k+r) \bmod n$. In this case,
\begin{equation}
    \label{eq:vqc}
    B(\theta) = \prod_{k=0}^{n-1} \cnot_{k, (k + r) \bmod n} \prod_{j=0}^{n-1} R(\alpha_j, \beta_j, \gamma_j),
\end{equation}
where $\theta$ denotes the set of parameters $(\alpha_j, \beta_j, \gamma_j)$ for $j = 0, \dots, n-1$. Applying $L$ layers of the unitary $B$ with (possibly different) parameter sets $\theta_1, \theta_2, \dots, \theta_L$, the complete entangling unitary is
\begin{equation*}
    U_\text{ent}(\theta) = B(\theta_L) B(\theta_{L-1}) \cdots B(\theta_1).
\end{equation*}
The resulting state $\ket{\psi_L(\theta)} = U_\text{ent}(\theta)\ket{\psi_{\bm{x}}}$ encodes the input node features into an entangled quantum state.

\begin{figure}[h]    
    \centering
    \includegraphics[width = \columnwidth]{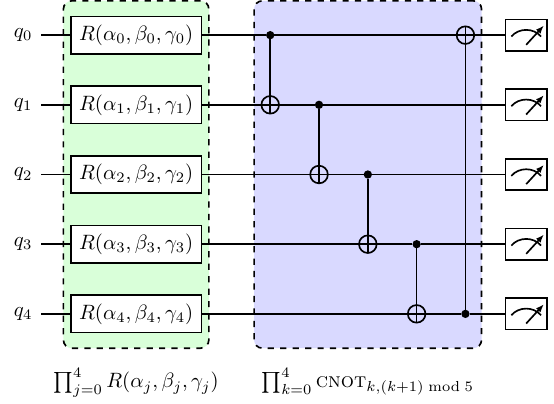}
    \caption{Five-qubit quantum feature extraction for a single layer of the unitary $B$ defined in \eqref{eq:vqc}.}
    \label{fig:vqc}
\end{figure}

Finally, the classical encoding of the node $\bm{x}$ is obtained by computing the Pauli-$Z$ expectation values on each qubit of $\ket{\psi_L(\theta)}$. Denoting the expectation value of qubit $i$ by $h_i$, the encoding is stored as the vector
\begin{equation}
    \label{eq:feature-enc}
    \bm{h} = \left(h_0, \dots, h_{n - 1} \right) \in \R^{n}.
\end{equation}
Figure~\ref{fig:vqc} shows a $5$-qubit entangling unitary circuit $U_\mathrm{ent}$ for a single layer ($L = 1$) and the final measurements.

\subsection{Quantum Message Passing}

In this section, we formulate a quantum analogue of classical message passing by adapting the alternating-operator structure of the Quantum Alternating Operator Ansatz (QAOA), a framework for constructing parameterized quantum circuits for combinatorial optimization and constraint-satisfaction problems. It was originally introduced by Farhi, Goldstone, and Gutmann~\cite{farhi2014quantum} as a quantum algorithm for approximate optimization, and later generalized by Hadfield et al.~\cite{hadfield2019quantum} into a flexible ansatz applicable to a broad class of problems. While QAOA is traditionally employed for global combinatorial optimization, its alternating ansatz of \textit{cost} and \textit{mixer} Hamiltonians provides a rigorous framework for inducing graph-structured state evolution. In the following, we first briefly review the QAOA framework, and then instantiate the general operators to obtain circuits that decompose into local operators and can be efficiently implemented.

\subsubsection{The QAOA framework}

At a high level, QAOA alternates between two families of unitary operators: \emph{problem} (or phase-separation) operators, which encode the objective function, and \emph{mixing} operators, which drive transitions between candidate solutions. This alternating structure defines both the expressivity and the algorithmic intuition behind QAOA.

Formally, a depth-$p$ QAOA circuit prepares a quantum state of the form
\[
    \ket{\psi(\bm{\gamma}, \bm{\beta})} = \prod_{k=1}^{p} \left( e^{-i \beta_k H_M} e^{-i \gamma_k H_C} \right) \ket{\psi_0},
\]
where $H_C$ is the \emph{problem} (or \text{cost}) Hamiltonian, typically diagonal in the computational basis and encoding the cost function, $H_M$ is the
\emph{mixing} Hamiltonian, $\ket{\psi_0}$ is an initial state (often the uniform superposition), and $\bm{\gamma}=(\gamma_1,\ldots,\gamma_p)$ and
% \begin{figure*}[h]
%     \centering
%     \includegraphics[width = \textwidth]{qaoa_general.pdf}
%     \caption{General QAOA circuit with input $\ket{\psi_0}$ and parameter sets $(\beta_1, \ldots, \beta_p)$ and $(\gamma_1,\ldots,\gamma_p)$.}
%     \label{fig:qaoa}
% \end{figure*}
$\bm{\beta}=(\beta_1,\ldots,\beta_p)$ are real parameters optimized by a classical outer loop.
% The general QAOA circuit is shown in Figure \ref{fig:qaoa}.

In the original formulation of QAOA for problems such as Max-Cut, the mixing Hamiltonian is chosen as $H_M = \sum_i X_i$, corresponding to global bit-flip operations, while $H_C$ is derived directly from the objective function of the instance. The generalized Quantum Alternating Operator Ansatz relaxes these choices by allowing problem-specific design of both operators. In particular, the mixing operator need not be a sum of Pauli-$X$ terms, and may be constructed to preserve hard constraints or to respect the structure of the feasible solution space.

From a conceptual perspective, QAOA can be interpreted in several complementary ways. It may be viewed as a digitized version of adiabatic quantum optimization, where increasing the depth $p$ yields progressively better approximations to continuous adiabatic evolution. It can also be understood as a variational quantum algorithm, or as a quantum analogue of classical local search or message-passing procedures, in which alternating operators play distinct and complementary roles.

A key practical feature of QAOA is that both the problem and mixing unitaries can often be decomposed into local quantum gates, such as single-qubit rotations and two-qubit controlled-phase operations. This locality makes QAOA particularly well-suited for implementation on near-term quantum hardware.

\subsubsection{An edge-local Quantum Message passing}
\label{sec:edge-lc}

We propose a quantum analogue of classical message passing inspired by QAOA.\footnote{Our message-passing circuit does not implement the global QAOA unitary, nor is it intended to. Our algorithm is local in nature and is only conceptually similar to the QAOA approach. The experiments in Section \ref{sec:expmnt} confirm that the algorithm is competitive.} Given the graph $G = (V,E)$, we define the cost Hamiltonian as
\begin{equation}
    H_C = \sum_{(u,v)\in E} \sum_{k=0}^{n-1} Z_{uk} Z_{vk},
\end{equation}
where $Z_{uk}$ denotes the Pauli-$Z$ operator acting on the $k$-th feature qubit associated with node $u$. The Hamiltonian $H_C$ induces pairwise interactions between neighboring nodes, acting feature-wise across each edge of the graph. The mixer Hamiltonian is defined as
\begin{equation}
    H_M = \sum_{u\in V} \sum_{k=0}^{n-1} X_{uk},
\end{equation}
where $X_{uk}$ denotes the Pauli-$X$ operator acting on the $k$-th feature qubit of node $u$. The mixer acts locally on each node’s feature register and redistributes amplitude among feature configurations. Alternating applications of the unitaries $e^{-i \gamma H_C}$ and $e^{-i \beta H_M}$ define a layered quantum evolution analogous to iterative message passing. The cost Hamiltonian $H_C$ encodes interactions between neighboring nodes by introducing feature-dependent phases, while the mixer Hamiltonian $H_M$ enables the propagation of this information through interference across the local feature space. Increasing the circuit depth corresponds to multiple rounds of quantum message passing on the underlying graph.

Hamiltonians of this form have appeared previously in the context of quantum graph optimization problems~\cite{hadfield2019quantum}. Here, we use the same type of feature-wise interaction to construct a quantum message-passing operation between neighboring nodes.

From the above Hamiltonians, the message-passing unitary corresponding to
a single QAOA layer with parameters $\beta$ and $\gamma$ is given by
\begin{align*}
    U_{\mathrm{MP}}(\gamma,\beta)
    & = e^{-i \beta H_M}\, e^{-i \gamma H_C} \\
    & = \prod_{u\in V} \prod_{k=0}^{n-1} R_{X_{uk}}(2\beta) \prod_{(u,v)\in E} \prod_{k=0}^{n-1} e^{-i \gamma Z_{uk} Z_{vk}},
\end{align*}
where $R_{X_{uk}}(\theta) = e^{-i \theta X_{uk}/2}$. The second equality follows from the fact that the operators $Z_{uk} Z_{vk}$ commute for different $(u,v)\in E$ and different feature indices $k$, and similarly the operators $X_{uk}$ commute for different nodes $u\in V$ and feature indices $k$.

Each component of the unitary $U_{\mathrm{MP}}$ can be efficiently implemented using standard quantum gates. The unitary $R_{X_{uk}}(2\beta)$ is a single-qubit rotation applied to the $k$-th feature qubit associated with node $u$.

The two-qubit unitaries $e^{-i \gamma Z_{uk} Z_{vk}}$ admit an efficient decomposition in terms of $\cnot{}$ and single-qubit rotation gates. Specifically, for each edge $(u,v) \in E$ and feature index $k$, we may write
\begin{equation}
    \label{eq:edge-uni}
    e^{-i \gamma Z_{uk} Z_{vk}} = \cnot_{uk \to vk} \cdot R_{Z_{vk}}(2\gamma) \cdot \cnot_{uk \to vk},
\end{equation}
where
\begin{itemize}
    \item $\cnot_{uk \to vk}$ denotes a $\cnot{}$ gate with control qubit $uk$ and target qubit $vk$.
    \item $R_{Z_{vk}}(2\gamma)$ is a single-qubit rotation applied to the feature qubit $k$ of the node $v$.
\end{itemize}
The circuit complexity of the unitary $U_{\mathrm{MP}}$ constructed from the above gates scales linearly with the number of input qubits. More precisely, the graph consists of $N$ nodes, each represented by an $n$-qubit feature register, so the total number of qubits required is $Nn$. Since each feature qubit participates in a constant number of single- and two-qubit gates per layer, the gate complexity of a single message-passing layer is $O(Nn)$.

For a quantum computer with sufficient qubit resources, this scaling is nearly optimal. However, the simultaneous use of a large number of qubits becomes prohibitive on NISQ hardware. To address this limitation, we next propose a qubit-efficient variant of $U_{\mathrm{MP}}$ that follows the same underlying message-passing logic while significantly reducing the number of qubits required.

\subsubsection{Edge-Local and Qubit-Efficient Quantum Message Passing}
\label{sec:edge-lc-qubit-efc}

\begin{figure}[h]
    \centering
    \includegraphics[width = \columnwidth]{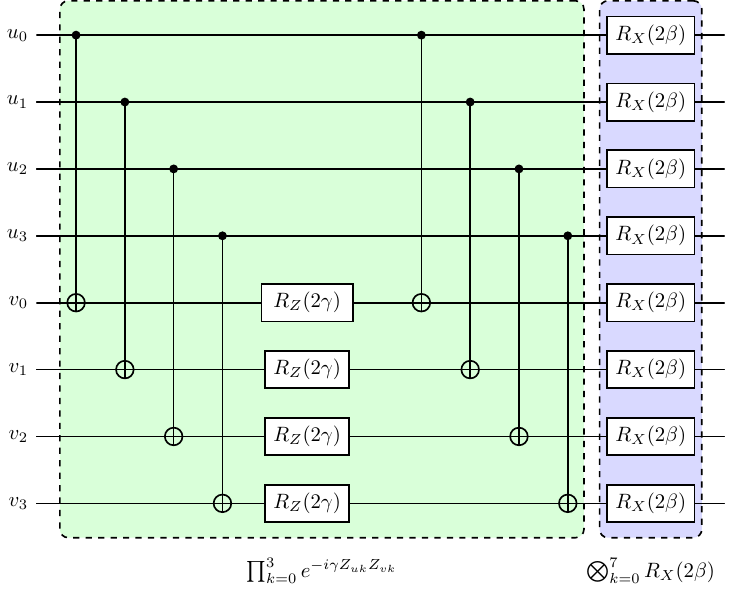}
    \caption{The quantum message-passing unitary $U_{\mathrm{MP},uv}(\gamma,\beta)$ for an edge $(u,v)$, where each node is represented by $n=4$ qubits.}
    \label{fig:qmp-local}
\end{figure}

Consider an edge $(u,v)\in E$ of the graph $G$. The action of the unitary $U_{\mathrm{MP}}$ restricted to this edge can be expressed as:
% \[ U_{\mathrm{MP},uv}(\gamma,\beta) := \prod_{k=0}^{\otimes 2n} R_{X_{uk}}(2\beta) \; \prod_{k=0}^{n-1} e^{-i \gamma Z_{uk} Z_{vk}}.\]
\[
U_{\mathrm{MP},uv}(\gamma,\beta)
:=
\bigotimes_{k=0}^{2n-1} R_{X_k}(2\beta)
\prod_{k=0}^{n-1} e^{-i\gamma Z_{u_k}Z_{v_k}}.
\]
This unitary can be efficiently implemented using the decomposition described in~\eqref{eq:edge-uni}. Figure~\ref{fig:qmp-local} illustrates the corresponding circuit for $U_{\mathrm{MP},uv}(\gamma,\beta)$ in the case where each node is represented by an $n=4$ qubit feature register.

The key idea behind our qubit-efficient construction is to apply the edge-local unitary $U_{\mathrm{MP},uv}$ separately for each edge $(u,v)\in E$, rather than implementing the full unitary $U_{\mathrm{MP}}$ on all nodes simultaneously. To obtain embeddings for all nodes in $G$, we first apply the quantum feature extraction procedure of Section~\ref{sec:qfe} to each node individually, producing an $n$-qubit feature encoding. We then iterate over the edges $(u,v)\in E$, and for each edge apply the unitary $U_{\mathrm{MP},uv}$ to the feature registers corresponding to nodes $u$ and $v$. Measurement of the output yields updated embeddings for both nodes.

Crucially, at no point does this procedure require more than $2n$ qubits simultaneously. The quantum feature extraction step uses $n$ qubits, as it operates on one node at a time, while each application of $U_{\mathrm{MP},uv}$ uses $2n$ qubits, acting only on the pair of nodes associated with a single edge.

For a concrete implementation of the above algorithm, we must address two practical questions:
\begin{enumerate}
    \item[(i)] After the quantum feature extraction step, each node is represented by an $n$-dimensional classical encoding. How should these encodings be mapped to quantum states for the quantum message-passing step?
    \item[(ii)] Since the quantum message-passing procedure iterates over edges, nodes with multiple neighbors will produce multiple embeddings. How can these be combined into a single embedding per node?
\end{enumerate}

Both problems admit natural and efficient solutions. We address problem~(i) using a standard data-encoding strategy from variational quantum algorithms, namely \emph{angle encoding}. Given a vector $\bm{h} = (h_0,h_1,\dots,h_{n-1}) \in \mathbb{R}^n$, we encode $\bm{h}$ into an $n$-qubit quantum state $\ket{\phi_{\bm{h}}}$ by applying single-qubit rotations to the computational basis state:

\begin{align*}
    \ket{\phi_{\bm{h}}}
    & = \bigotimes_{k = 0}^{n - 1} R_Y(2h_k) \ket{0}^{\otimes n} \\
    & = \bigotimes_{k = 0}^{n - 1} \big(\cos(h_k)\ket{0} + \sin(h_k)\ket{1} \big)
\end{align*}
Therefore, during the message-passing step, for each edge $(u,v)\in E$ with classical encodings $\bm{h}_u$ and $\bm{h}_v$, we prepare the states $\ket{\phi_{\bm{h}_u}}$ and $\ket{\phi_{\bm{h}_v}}$ and apply the edge-local quantum message-passing circuit shown in Figure~\ref{fig:qmp-local}.

For the $k$-th feature coordinate, the expectation value obtained
from the register associated with node $u$ is

\begin{align*}
    \langle Z_{u,k}\rangle_{\mathrm{out}}
    =
    &\cos(2\beta)\cos(2h_{u,k}) \\
    &+
    \sin(2\beta)\sin(2\gamma)
    \sin(2h_{u,k})\cos(2h_{v,k}).
\end{align*}

where $\bm{h}_{u, k}$ is the $k$-th coordinate of $\bm{h}_u$. The second term depends on both endpoints of the edge. Therefore, when $\gamma\neq0$, the measured message associated with node $u$ is sensitive to the neighboring feature $\bm{h}_{v,k}$. When $\gamma=0$, this term disappears.

Problem~(ii) is handled by a simple aggregation procedure. For nodes with multiple neighbors, the message-passing step yields multiple embeddings, one from each incident edge. A unique embedding for each node is obtained by aggregating these embeddings, for example by summation or averaging. Nodes with no neighbors do not participate in the message-passing step, and their embeddings are given directly by the output of the quantum feature extraction procedure.

Algorithm~\ref{alg:graph-embd} summarizes the full procedure for computing node embeddings.

\begin{algorithm}
    \caption{Graph embedding}
    \label{alg:graph-embd}
    \begin{algorithmic}[1]
        \Require Graph $G = (V, E)$ with $\abs{V} = N$, feature matrix $X \in \R^{N \times d}$
        \Ensure An embedding for each node of $G$

        \State Initialize the list of embeddings $\mathcal{E}$.
        \State Run the quantum feature extraction to obtain the encodings $\bm{h}_{v_0}, \dots, \bm{h}_{v_{N - 1}}$ for the nodes in $G$.
        \For {$(u, v) \in E$}
            \State Prepare the angle encodings $\ket{\phi_{\bm{h}_u}}$ and $\ket{\phi_{\bm{h}_v}}$.
            \State \parbox[t]{\dimexpr\linewidth-\algorithmicindent\relax}{Run the quantum message passing on the states $\ket{\phi_{\bm{h}_u}}$, $\ket{\phi_{\bm{h}_v}}$.} \vspace*{0.3mm}
            \State \parbox[t]{\dimexpr\linewidth-\algorithmicindent\relax}{Measure the resulting output to obtain embeddings for $u$ and $v$.} \vspace*{0.3mm}
            \State Add the embeddings to $\mathcal{E}$.
        \EndFor
        \State Refine $\mathcal{E}$ by replacing nodes with multiple embeddings with their sum.
    \end{algorithmic}
\end{algorithm}

\subsubsection{Trade-off Analysis}

The crucial distinction between the edge-local algorithm of Section~\ref{sec:edge-lc} and the qubit-efficient version proposed in Section~\ref{sec:edge-lc-qubit-efc} lies in the trade-off between qubit resources and encoding redundancy.

In the edge-local algorithm of Section~\ref{sec:edge-lc}, each node is represented by an $n$-qubit feature register, requiring a total of $Nn$ qubits for a graph with $N$ nodes. This approach allows each node to be encoded at most once, requiring $O(N)$ angle encodings. However, the hardware requirements scale linearly with the size of the graph, which becomes prohibitive for large graphs on near-term quantum devices.

In contrast, the qubit-efficient edge-local algorithm of Section~\ref{sec:edge-lc-qubit-efc} constructs a small circuit for each edge $(u,v)$ using only $2n$ qubits. This reduces the spatial (qubit) complexity from $O(Nn)$ to $O(n)$, rendering the approach feasible on current quantum hardware regardless of graph size. The trade-off is an increase in encoding redundancy: since a node $u$ participates in $\deg(u)$ edges, its feature vector $\bm{h}_u$ must be re-encoded for each incident edge.

Consequently, the total number of angle encoding operations scales as
\[ \sum_{u\in V} \deg(u) = 2|E|. \]
For sparse graphs, this increase does not significantly affect the overall computational cost compared to the algorithm of Section~\ref{sec:edge-lc}. In particular, if the maximum degree of the graph is bounded by a constant, then $|E| \in O(N)$, and both algorithms require $O(N)$ angle encodings.

\subsection{Deep Graph Infomax Objective}

To learn unsupervised node representations, we employ the Deep Graph Infomax (DGI) objective~\cite{velickovic2019deep}, which maximizes a contrastive lower bound on the mutual information between local node representations and a global summary of the graph.

Let $\mathcal{H}=\{\bm{h}_0,\dots,\bm{h}_{N-1}\}$, where $\bm{h}_j\in\mathbb{R}^n$, denote the node embeddings produced by Algorithm~\ref{alg:graph-embd}. We construct the global summary vector by taking the average of the node embeddings and applying a sigmoid:
\[
    \bm{s}
    =
    \sigma\left(
    \frac{1}{N}
    \sum_{j=0}^{N-1}\bm{h}_j
    \right).
\]

The DGI framework employs a discriminator $D(\bm{h},\bm{s})$ that assigns high scores to ``true'' node-summary pairs and low scores to ``false'' pairs obtained via a corruption function. Positive samples are given by $(\bm{h}_j,\bm{s})$ for $j\in\{0,\dots,N-1\}$. Negative samples are generated by applying a corruption operator (in our case, a permutation of the input features) to the graph $G$, and then re-running Algorithm~\ref{alg:graph-embd} to produce corrupted embeddings $\mathcal{H}'=\{\bm{h}_0',\dots,\bm{h}_{N-1}'\}$. These are paired with the original summary vector to form the pairs $(\bm{h}_j',\bm{s})$.

We optimize a binary cross-entropy objective over $N$ positive and $N$ negative pairs:
\[
    \mathcal{L}_{\mathrm{DGI}} = -\frac{1}{2N}\sum_{j=0}^{N-1} \Big( \log D(\bm{h}_j,\bm{s}) + \log\big(1 - D(\bm{h}_j',\bm{s})\big) \Big).
\]
By optimizing $\mathcal{L}_{\mathrm{DGI}}$, the encoder learns node representations that are predictive of the global graph context, encouraging $\mathcal{H}$ to capture both local structural information and global contextual information.

\subsubsection{Complexity analysis}

In the following, we provide a brief comparison of the running-time complexity of our algorithm with its classical counterpart. We also include the complexity of a \emph{hybrid} algorithm implemented for our experiments (Section~\ref{sec:expmnt}), which uses the same quantum feature extraction layer as our model but replaces quantum message passing with a classical message-passing procedure. Throughout this analysis, we assume the number of layers $L$ is a constant, consistent with our implementation. For classical algorithms, we count the number of elementary arithmetic operations, while for quantum algorithms we count the number of quantum gates.

\textbf{Classical GCN.}
In a classical GCN, Equation~\eqref{eq:classical-gcn}, the feature transformation step computes the matrix product $H^{(l)}W^{(l)}$, where $H^{(l)} \in \mathbb{R}^{N \times d}$ and $W^{(l)} \in \mathbb{R}^{d \times d}$. This requires $O(Nd^2)$ operations. The message-passing step multiplies by the normalized adjacency matrix $D^{-1/2} \widetilde{A} D^{-1/2}$, which can be computed in $O(|E|d)$ operations under the assumption that $\widetilde{A}$ is sparse. The activation function $\sigma$ is applied componentwise and contributes $O(Nd)$ operations. Thus, the overall complexity of a classical GCN layer is $O(Nd^2 + |E|d)$.

\textbf{Quantum algorithm.}
In our quantum algorithm (Algorithm~\ref{alg:graph-embd}), the initial amplitude encoding of classical feature vectors into quantum states requires $O(Nd)$ single-qubit rotations.\footnote{While the cost of classical-to-quantum data encoding is often ignored in the literature, we explicitly include it here for a fair comparison with classical algorithms.} The quantum feature extraction step applies a fixed quantum circuit of size $O(\log d)$ to each node. Since this circuit is applied sequentially to all $N$ nodes, the total gate complexity of this step is $O(N\log d)$. The quantum message-passing step iterates over the edges of the graph and applies an edge-local circuit of size $O(\log d)$ per edge, resulting in a total complexity of $O(|E|\log d)$. Combining these contributions, the overall complexity of Algorithm~\ref{alg:graph-embd} is $O(Nd + |E|\log d)$.

\textbf{Hybrid algorithm.}
In the hybrid algorithm, the initial amplitude encoding and quantum feature extraction steps have complexities $O(Nd)$ and $O(N\log d)$, respectively. The subsequent message-passing step is performed classically and requires $O(|E|\log d)$ operations. Consequently, the overall complexity of the hybrid algorithm is also $O(Nd + |E|\log d)$, matching our proposed algorithm.

\section{Experiments}
\label{sec:expmnt}

Our experiments are conducted on two datasets: the Cora citation network and the Phase~3 release of the 1000 Genomes Project. Both datasets provide external labels, but these labels are completely withheld during graph construction and model training. They are only loaded after training to evaluate the final node embeddings. This allows us to evaluate the agreement between the learned representation and an external class structure without using the labels to optimize the model.

\paragraph{1000 Genomes Phase 3 SNP Dataset.}
The original dataset contains $N=2504$ individuals. Each individual is represented by diploid SNP genotypes, while the official sample panel provides population and super-population annotations \cite{10002015global}. In this work, we consider the five super-populations
\[
\{\mathrm{AFR},\mathrm{AMR},\mathrm{EAS},\mathrm{EUR},\mathrm{SAS}\}
\]
as the external reference labels. The 26 finer population labels are not considered in the present experiments.

To reduce the effect of correlated genomic regions, linkage disequilibrium pruning (LD-pruning) \cite{chang2015second} is performed using a $500$ kb window, a step size of one variant, and an $r^2$ threshold of $0.2$. We then remove one individual from each second-degree-or-closer related pair using a KING kinship \cite{manichaikul2010robust} cutoff of $0.0884$. This leaves $N=2467$ individuals for the remaining experiments. We then divide the genome into two non-overlapping views. SNPs from odd-numbered chromosomes are used to construct the graph, while SNPs from even-numbered chromosomes are used to construct the node features. This prevents the graph and the node features from being two transformations of exactly the same SNP subset.

For the graph view, principal component analysis is applied to the LD-pruned odd-chromosome SNPs, and the first $32$ components are retained. Let $\bm{g}_i \in \R^{32}$ denote the resulting graph-construction representation of individual $i$. Each coordinate is standardized across individuals, and every vector is subsequently normalized to unit Euclidean norm. The cosine similarity between two individuals is then defined as
\[
S_{ij}
=
\frac{\lrang{\bm{g}_i,\bm{g}_j}}
{\opnorm{\bm{g}_i}\opnorm{\bm{g}_j}},
\]
where $\lrang{\bm{g}_i,\bm{g}_j}$ is the Euclidean inner product and $\opnorm{\cdot}$ is the Euclidean norm.

For each individual $i$, we identify the set of its $k=10$ nearest neighbors, denoted $\mathcal{N}_{10}(i)$. We use a symmetric union $k$-nearest-neighbor graph, where an undirected edge is included when either endpoint selects the other:
\[
(i,j)\in E
\iff
j\in\mathcal{N}_{10}(i)
\ \lor\
i\in\mathcal{N}_{10}(j).
\]
This construction yields $16\,568$ undirected edges, corresponding to $33\,136$ directed edges in the PyTorch Geometric representation. The resulting graph contains three connected components, with $2417$ individuals, or approximately $97.97\%$ of all nodes, belonging to the largest component. The graph has no isolated nodes, a minimum degree of $10$, and an average degree of approximately $13.43$.

For the node-feature view, PCA is separately applied to the LD-pruned even-chromosome SNPs. The first $64$ components are retained, standardized, and normalized to unit length. Each node is therefore associated with $\bm{x}_i \in \R^{64}$. Since $64=2^6$, the amplitude-encoding feature extractor uses six qubits without requiring additional zero padding. The local quantum message-passing circuit uses two six-qubit registers and therefore contains $12$ wires.

\paragraph{Cora Citation Network.}
The dataset, which is already presented as a graph, represents academic papers as nodes, each labeled with one of seven research categories, while edges correspond to citation links between papers. Each node is associated with a feature vector given by a bag-of-words (BoW) representation of the paper's abstract. The dataset contains $2708$ nodes, $5429$ edges, $1433$ features, and seven distinct classes. We use the Cora citation network from the Planetoid benchmark
\cite{yang2016revisiting}.

For the Cora dataset, $\bm{x}_i \in \R^{1433}$. The amplitude-encoding circuit therefore uses $\left\lceil \log_2(1433) \right\rceil = 11$ qubits, and the input is padded to dimension $2^{11}=2048$. The local message-passing circuit contains two $11$-qubit registers and therefore uses $22$ quantum wires. The original unweighted citation graph is used without adding feature-similarity edge weights.

All neural models are trained in a full-graph setting using the Adam optimizer. The complete procedure is repeated over the random seeds $\{0,1,\ldots,9\}$. A new model and optimizer are initialized for every neural model and every seed. The DGI corruption, parameter initialization, and final $k$-means clustering are controlled by the corresponding seed. The same set of seeds is also used for the clustering baselines.

\subsection{Benchmark Models}
\label{sec:benchmarks}

We include classical, graph-only, hybrid, and quantum baselines.

The \emph{$k$-means} baseline measures the structure that is already present in the node features. For the genomic dataset, the prepared $64$ even-chromosome principal components are directly clustered. For Cora, $k$-means is applied directly to the original $1433$-dimensional BoW node features.

The \emph{spectral clustering} baseline measures the structure that is already present in the graph. For Cora, every citation edge is assigned unit weight. For the genomic dataset, the cosine-similarity values of the constructed $k$-nearest-neighbor graph are used as the affinity weights.

The \emph{GCN + DGI} baseline replaces the quantum encoder with a classical graph convolutional network while retaining the same DGI objective. We report two configurations because there are two reasonable choices for the embedding dimension. In the first configuration, we set the embedding dimension equal to the number of qubits used by our quantum model: $11$ for Cora and $6$ for the 1000 Genomes dataset. This gives a dimension-matched comparison, so the classical and quantum models are evaluated with the same output size. In the second configuration, we use an embedding dimension of $512$, following the setting used in the original DGI implementation~\cite{velickovic2019deep}. We include this second configuration because $512$ dimensions represent the standard setting of the classical baseline and show its performance without restricting its embedding size to match the quantum model. Therefore, the two configurations answer two different questions: the dimension-matched setting compares the models under the same output dimension, while the $512$-dimensional setting compares our model with the standard GCN + DGI baseline.

The hybrid baseline uses the same quantum feature extractor as the proposed model but replaces the quantum message-passing operation with classical neighborhood aggregation. This comparison separates the effect of quantum feature extraction from the effect of the quantum graph interaction.

The $\gamma=0$ experiment uses our proposed architecture while fixing all $ZZ$ interaction parameters to zero. This removes the interaction-mediated dependence on the neighboring register while retaining the remaining circuit structure.

\subsection{Evaluation Metrics}
\label{sec:evaluationmetrics}

For methods that produce learned node embeddings, $k$-means is independently applied to the embedding obtained from each seed. The $k$-means baseline is applied directly to the input node features, while spectral clustering directly produces the cluster assignments. We use $K=7$ for Cora and $K=5$ for the genomic dataset. The number of clusters is fixed according to the number of external classes and is not selected by maximizing an evaluation metric.

The primary evaluation metric is Normalized Mutual Information. Let $\widehat{\bm{c}}=(\widehat{c}_1,\ldots,\widehat{c}_N)$ denote the predicted cluster assignments and let $\bm{y}=(y_1,\ldots,y_N)$ denote the external labels. We use the arithmetic normalization
\[
    \operatorname{NMI}(\widehat{\bm{c}},\bm{y})
    =
    \frac{2I(\widehat{\bm{c}};\bm{y})}
    {H(\widehat{\bm{c}})+H(\bm{y})}.
\]
Here, $I(\widehat{\bm{c}};\bm{y})$ denotes the mutual information between the predicted cluster assignments and the external labels, while $H(\widehat{\bm{c}})$ and $H(\bm{y})$ denote their Shannon entropies. NMI is invariant to a permutation of the cluster identifiers and takes a value close to one when the predicted clustering closely agrees with the external labels.

We additionally report the Adjusted Rand Index and Adjusted Mutual Information. ARI evaluates pairwise agreement between the predicted and external partitions while correcting for agreement expected by chance. AMI similarly adjusts the mutual information for agreement expected by chance.

Since cluster identifiers do not have a predefined correspondence with class identifiers, accuracy cannot be calculated directly from the raw cluster assignments. We therefore report matched clustering accuracy. Let $\Pi_K$ denote the set of all one-to-one permutations of the $K$ cluster identifiers. The matched accuracy is defined as
\[
\operatorname{Acc}_{\mathrm{match}}
=
\frac{1}{N}
\max_{\pi\in\Pi_K}
\sum_{i=1}^{N}
\mathbb{I}
\left[
y_i=\pi(\widehat{c}_i)
\right].
\]
Here, $N$ is the number of nodes, $y_i$ is the external label of node $i$, $\widehat{c}_i$ is its predicted cluster identifier, $\pi$ maps the predicted cluster identifiers to the external class identifiers, and $\mathbb{I}[\cdot]$ is equal to one when its condition is true and zero otherwise. The maximizing permutation is obtained using the Hungarian assignment algorithm.

\subsection{Results}
\label{sec:results}

\begin{table*}[h]
    \centering
    \caption{Clustering performance on the Cora dataset.}
    \label{tab:cora_clustering_benchmarks}

    \renewcommand{\arraystretch}{1.25}
    \setlength{\tabcolsep}{3.8pt}
    \footnotesize

    \begin{tabular*}{\textwidth}{@{\extracolsep{\fill}}lcccc@{}}
        \toprule
        \textbf{Model}
        &
        \textbf{NMI}
        &
        \textbf{ARI}
        &
        \textbf{AMI}
        &
        \textbf{Matched Acc.}
        \\
        \midrule

        $k$-means
        &
        $0.1202\pm0.0319$
        &
        $0.0511\pm0.0215$
        &
        $0.1165\pm0.0320$
        &
        $0.3162\pm0.0321$
        \\

        Spectral clustering
        &
        $0.0221\pm0.0085$
        &
        $-0.0044\pm0.0032$
        &
        $0.0151\pm0.0086$
        &
        $0.2942\pm0.0054$
        \\

        Classical GCN + DGI
        (embedding-dimension matched)
        &
        $0.1974\pm0.0573$
        &
        $0.1221\pm0.0496$
        &
        $0.1945\pm0.0575$
        &
        $0.3370\pm0.0620$
        \\

        Classical GCN + DGI
        (embedding dimension $512$)
        &
        $0.5645\pm0.0075$
        &
        $0.5140\pm0.0232$
        &
        $0.5629\pm0.0075$
        &
        $0.7167\pm0.0188$
        \\

        Hybrid Model + DGI
        &
        $0.1854\pm0.0267$
        &
        $0.1260\pm0.0874$
        &
        $0.1833\pm0.0185$
        &
        $0.3287\pm0.0543$
        \\

        Our model, $\gamma=0$
        &
        $0.1293\pm0.0115$
        &
        $0.0654\pm0.0298$
        &
        $0.1289\pm0.0243$
        &
        $0.2961\pm0.0114$
        \\

        Our model
        &
        $0.3906\pm0.0705$
        &
        $0.3183\pm0.0634$
        &
        $0.3829\pm0.0825$
        &
        $0.5986\pm0.0969$
        \\

        \bottomrule
    \end{tabular*}
\end{table*}

\begin{table*}[h]
    \centering
    \caption{Clustering performance on the 1000 Genomes Phase~3 dataset.}
    \label{tab:genomic_clustering_benchmarks}

    \renewcommand{\arraystretch}{1.25}
    \setlength{\tabcolsep}{3.8pt}
    \footnotesize

    \begin{tabular*}{\textwidth}{@{\extracolsep{\fill}}lcccc@{}}
        \toprule
        \textbf{Model}
        &
        \textbf{NMI}
        &
        \textbf{ARI}
        &
        \textbf{AMI}
        &
        \textbf{Matched Acc.}
        \\
        \midrule

        $k$-means
        &
        $0.5414\pm0.0522$
        &
        $0.3028\pm0.0795$
        &
        $0.5403\pm0.0523$
        &
        $0.5927\pm0.0486$
        \\

        Spectral clustering
        &
        $0.6651\pm0.0137$
        &
        $0.4211\pm0.0149$
        &
        $0.6642\pm0.0137$
        &
        $0.6071\pm0.0085$
        \\

        Classical GCN + DGI
        (embedding-dimension matched)
        &
        $0.2337\pm0.0917$
        &
        $0.1678\pm0.0893$
        &
        $0.2322\pm0.0919$
        &
        $0.4282\pm0.0762$
        \\

        Classical GCN + DGI
        (embedding dimension $512$)
        &
        $0.4960\pm0.0597$
        &
        $0.3203\pm0.0588$
        &
        $0.4949\pm0.0598$
        &
        $0.5760\pm0.0760$
        \\

        Hybrid Model + DGI
        &
        $0.2846\pm0.0151$
        &
        $0.2107\pm0.0374$
        &
        $0.2832\pm0.0296$
        &
        $0.4564\pm0.0379$
        \\

        Our model, $\gamma=0$
        &
        $0.1846\pm0.0519$
        &
        $0.0984\pm0.0239$
        &
        $0.1796\pm0.0173$
        &
        $0.3069\pm0.0192$
        \\

        Our model
        &
        $0.5183\pm0.0837$
        &
        $0.2760\pm0.0749$
        &
        $0.5092\pm0.0478$
        &
        $0.5823\pm0.0902$
        \\

        \bottomrule
    \end{tabular*}

\end{table*}

\begin{figure}
    \centering

    \includegraphics[width=0.98\columnwidth]{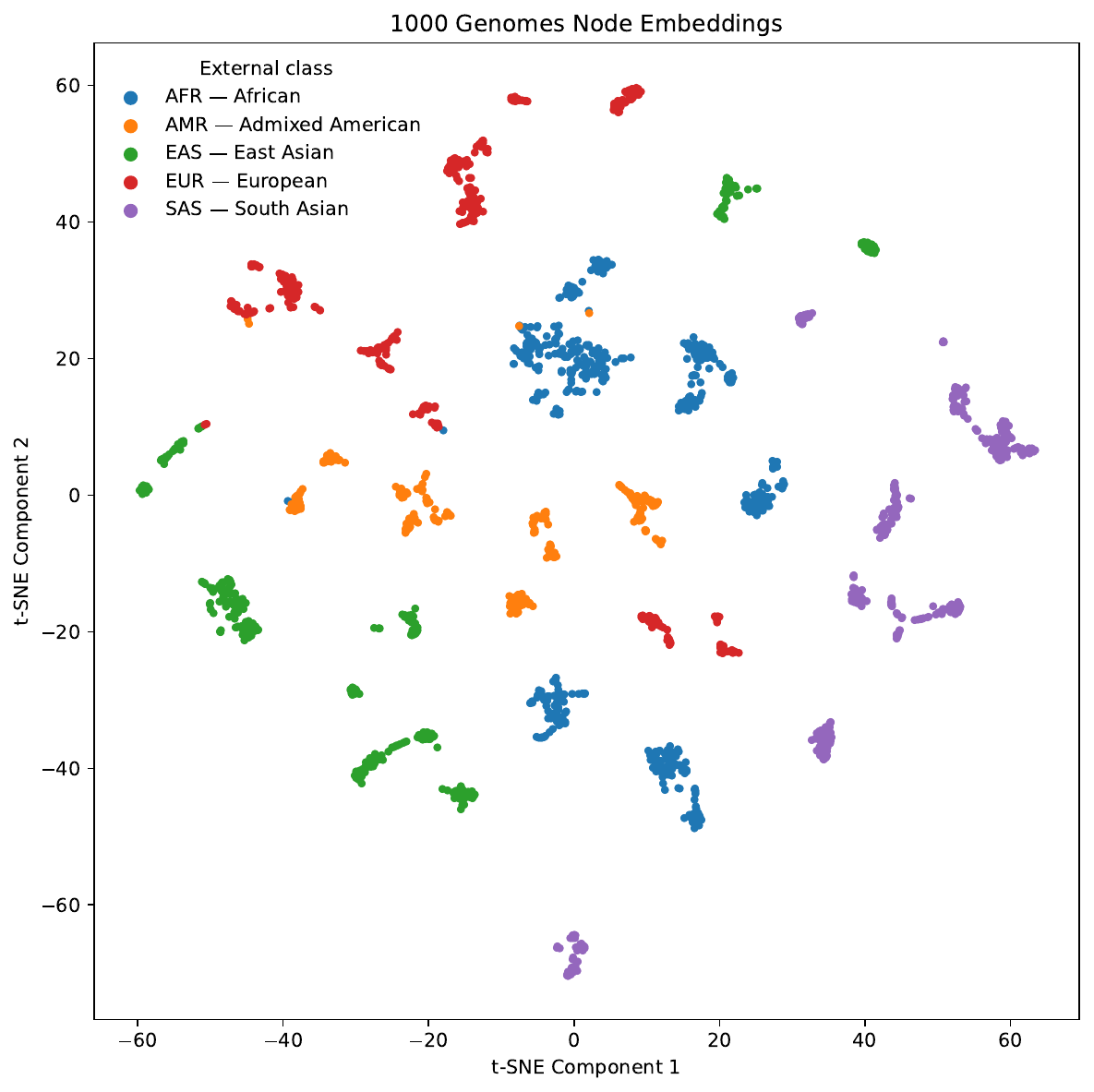}
    \caption{
    The t-SNE visualization of the node representations learned on the 1000 Genomes Phase~3 dataset. Colors indicate the five external super-population annotations.
    }
    \label{fig:snp}
\end{figure}

\begin{figure}
    \centering

    \includegraphics[width=0.98\columnwidth]{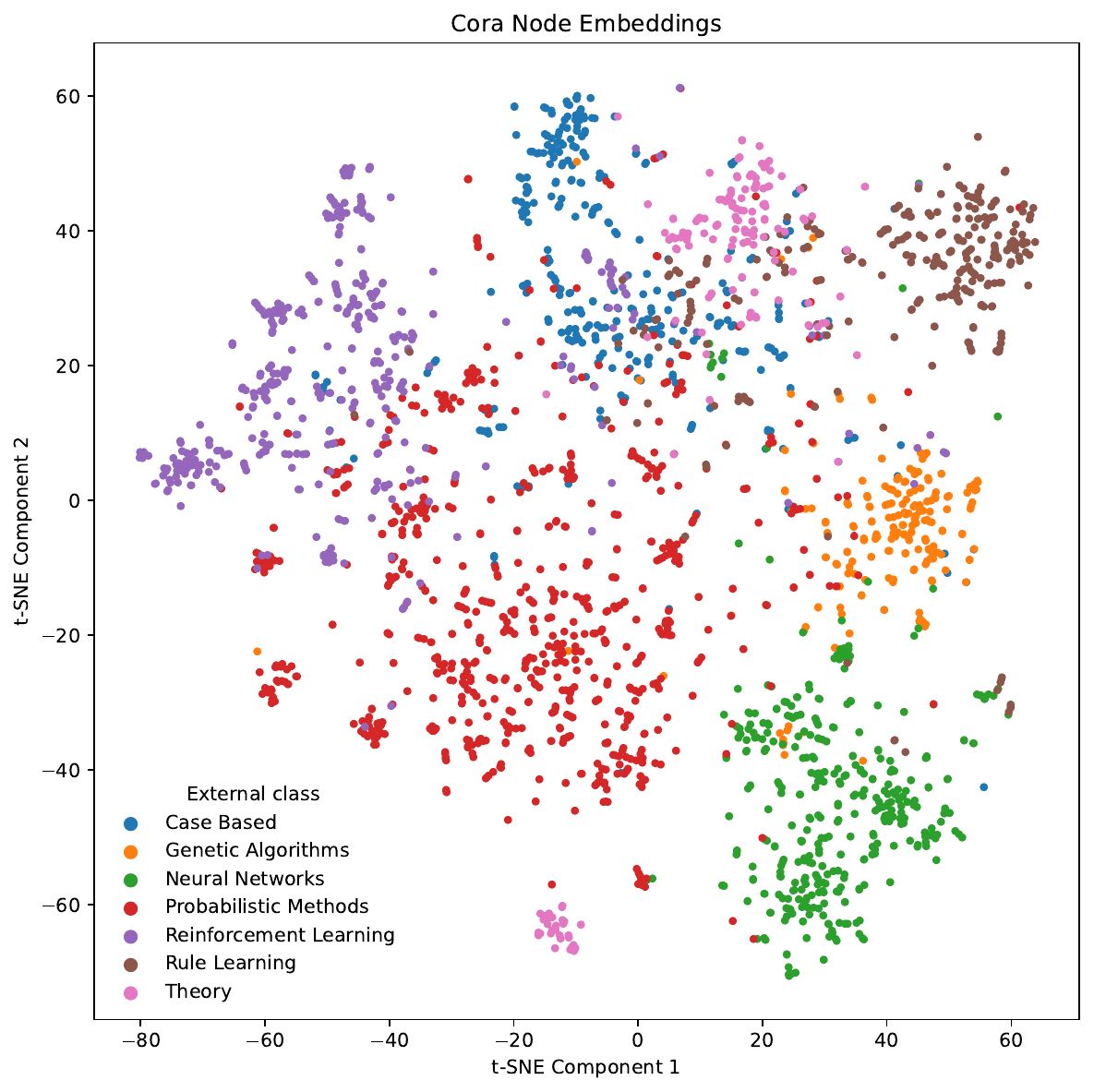}
    \caption{
    The t-SNE visualization of the node representations learned on the Cora citation network. Colors indicate the seven external research categories. The displayed run is from the seed whose NMI is closest to the ten-seed mean.
    }
    \label{fig:cora}
\end{figure}

Figure~\ref{fig:snp} shows a two-dimensional visualization of the genomic node embeddings, with colors indicating the five external super-population labels. Figure~\ref{fig:cora} shows the corresponding visualization for Cora. To avoid selecting a visually favorable random seed, the displayed run for each dataset is chosen as the seed whose NMI is closest to the ten-seed mean.

The clustering results for the Cora citation network and the 1000 Genomes Phase~3 dataset are shown in Table~\ref{tab:cora_clustering_benchmarks} and Table~\ref{tab:genomic_clustering_benchmarks}, respectively. All values are reported as mean $\pm$ sample standard deviation over ten random seeds.

On Cora, our model has higher mean NMI, ARI, AMI, and matched accuracy than the dimension-matched GCN + DGI baseline, the hybrid model, and the $\gamma=0$ experiment. The $512$-dimensional GCN + DGI configuration remains the strongest baseline on this dataset.

For the 1000 Genomes dataset, both the $k$-means and spectral clustering baselines show that class structure is already present in the node features and in the graph. Spectral clustering gives the highest mean values on this dataset. Our model again has higher mean values than the dimension-matched GCN + DGI baseline, the hybrid model, and the $\gamma=0$ experiment for all four metrics. Compared with the $512$-dimensional GCN + DGI configuration, our model has higher mean NMI, AMI, and matched accuracy, while the GCN + DGI baseline has a higher mean ARI.

The difference between the full model and the $\gamma=0$ experiment is consistent across both datasets. Since the remaining circuit structure is unchanged in this comparison, the decrease obtained when the $ZZ$ interaction is removed indicates the contribution of the interaction-dependent message term to the learned representation.

These experiments evaluate the quality of the learned unsupervised node representations. They do not establish a quantum computational advantage.

\newpage
\bibliographystyle{plain}
\bibliography{references}

\begin{thebibliography}{10}

\bibitem{ai2024quantum}
Xing Ai, Zhihong Zhang, Luzhe Sun, Junchi Yan, and Edwin Hancock.
\newblock Towards quantum graph neural networks: An ego-graph learning
  approach.
\newblock {\em arXiv preprint arXiv:2201.05158}, 2022.

\bibitem{barenco1995elementary}
Adriano Barenco, Charles~H Bennett, Richard Cleve, David~P DiVincenzo, Norman
  Margolus, Peter Shor, Tycho Sleator, John~A Smolin, and Harald Weinfurter.
\newblock Elementary gates for quantum computation.
\newblock {\em Physical review A}, 52(5):3457, 1995.

\bibitem{bharti2022noisy}
Kishor Bharti, Alba Cervera-Lierta, Thi~Ha Kyaw, Tobias Haug, Sumner
  Alperin-Lea, Abhinav Anand, Matthias Degroote, Hermanni Heimonen, Jakob~S
  Kottmann, Tim Menke, et~al.
\newblock Noisy intermediate-scale quantum algorithms.
\newblock {\em Reviews of Modern Physics}, 94(1):015004, 2022.

\bibitem{biamonte2017quantum}
Jacob Biamonte, Peter Wittek, Nicola Pancotti, Patrick Rebentrost, Nathan
  Wiebe, and Seth Lloyd.
\newblock Quantum machine learning.
\newblock {\em Nature}, 549(7671):195--202, 2017.

\bibitem{cerezo2021variational}
Marco Cerezo, Andrew Arrasmith, Ryan Babbush, Simon~C Benjamin, Suguru Endo,
  Keisuke Fujii, Jarrod~R McClean, Kosuke Mitarai, Xiao Yuan, Lukasz Cincio,
  et~al.
\newblock Variational quantum algorithms.
\newblock {\em Nature Reviews Physics}, 3(9):625--644, 2021.

\bibitem{chang2015second}
Christopher~C Chang, Carson~C Chow, Laurent~CAM Tellier, Shashaank Vattikuti,
  Shaun~M Purcell, and James~J Lee.
\newblock Second-generation plink: rising to the challenge of larger and richer
  datasets.
\newblock {\em Gigascience}, 4(1):s13742--015, 2015.

\bibitem{chen2021hybrid}
Samuel Yen-Chi Chen, Tzu-Chieh Wei, Chao Zhang, Haiwang Yu, and Shinjae Yoo.
\newblock Hybrid quantum-classical graph convolutional network.
\newblock {\em arXiv preprint arXiv:2101.06189}, 2021.

\bibitem{chen2025jet}
Yi-An Chen and Kai-Feng Chen.
\newblock Jet discrimination with a quantum complete graph neural network.
\newblock {\em Physical Review D}, 111(1):016020, 2025.

\bibitem{10002015global}
1000 Genomes~Project Consortium et~al.
\newblock A global reference for human genetic variation.
\newblock {\em Nature}, 526(7571):68, 2015.

\bibitem{farhi2014quantum}
Edward Farhi, Jeffrey Goldstone, and Sam Gutmann.
\newblock A quantum approximate optimization algorithm.
\newblock {\em arXiv preprint arXiv:1411.4028}, 2014.

\bibitem{farhi2018classification}
Edward Farhi and Hartmut Neven.
\newblock Classification with quantum neural networks on near term processors.
\newblock {\em arXiv preprint arXiv:1802.06002}, 2018.

\bibitem{gao2018large}
Hongyang Gao, Zhengyang Wang, and Shuiwang Ji.
\newblock Large-scale learnable graph convolutional networks.
\newblock pages 1416--1424, 2018.

\bibitem{gilmer2017neural}
Justin Gilmer, Samuel~S Schoenholz, Patrick~F Riley, Oriol Vinyals, and
  George~E Dahl.
\newblock Neural message passing for quantum chemistry.
\newblock In {\em International conference on machine learning}, pages
  1263--1272. Pmlr, 2017.

\bibitem{hadfield2019quantum}
Stuart Hadfield, Zhihui Wang, Bryan O’gorman, Eleanor~G Rieffel, Davide
  Venturelli, and Rupak Biswas.
\newblock From the quantum approximate optimization algorithm to a quantum
  alternating operator ansatz.
\newblock {\em Algorithms}, 12(2):34, 2019.

\bibitem{hamilton2017inductive}
Will Hamilton, Zhitao Ying, and Jure Leskovec.
\newblock Inductive representation learning on large graphs.
\newblock {\em Advances in neural information processing systems}, 30, 2017.

\bibitem{havlivcek2019supervised}
Vojt{\v{e}}ch Havl{\'\i}{\v{c}}ek, Antonio~D C{\'o}rcoles, Kristan Temme,
  Aram~W Harrow, Abhinav Kandala, Jerry~M Chow, and Jay~M Gambetta.
\newblock Supervised learning with quantum-enhanced feature spaces.
\newblock {\em Nature}, 567(7747):209--212, 2019.

\bibitem{wang2023quantumgcn}
Zhirui Hu, Jinyang Li, Zhenyu Pan, Shanglin Zhou, Lei Yang, Caiwen Ding, Omer
  Khan, Tong Geng, and Weiwen Jiang.
\newblock On the design of quantum graph convolutional neural network in the
  nisq-era and beyond.
\newblock pages 290--297, 2022.

\bibitem{kipf2016semi}
Thomas~N Kipf and Max Welling.
\newblock Semi-supervised classification with graph convolutional networks.
\newblock {\em arXiv preprint arXiv:1609.02907}, 2016.

\bibitem{larose2020robust}
Ryan LaRose and Brian Coyle.
\newblock Robust data encodings for quantum classifiers.
\newblock {\em arXiv preprint arXiv:2003.01695}, 2020.

\bibitem{manichaikul2010robust}
Ani Manichaikul, Josyf~C Mychaleckyj, Stephen~S Rich, Kathy Daly, Mich{\`e}le
  Sale, and Wei-Min Chen.
\newblock Robust relationship inference in genome-wide association studies.
\newblock {\em Bioinformatics}, 26(22):2867--2873, 2010.

\bibitem{mernyei2022equivariant}
P{\'e}ter Mernyei, Konstantinos Meichanetzidis, and Ismail~Ilkan Ceylan.
\newblock Equivariant quantum graph circuits.
\newblock In {\em International conference on machine learning}, pages
  15401--15420. PMLR, 2022.

\bibitem{preskill2018quantum}
John Preskill.
\newblock Quantum computing in the nisq era and beyond.
\newblock {\em Quantum}, 2:79, 2018.

\bibitem{salamatov2025quantum}
Azamat Salamatov and Gowtham Atluri.
\newblock Quantum and classical graph convolutional neural networks for protein
  ligand dissociation constant prediction.
\newblock {\em bioRxiv}, pages 2025--11, 2025.

\bibitem{scarselli2008graph}
Franco Scarselli, Marco Gori, Ah~Chung Tsoi, Markus Hagenbuchner, and Gabriele
  Monfardini.
\newblock The graph neural network model.
\newblock {\em IEEE transactions on neural networks}, 20(1):61--80, 2008.

\bibitem{schlichtkrull2018modeling}
Michael Schlichtkrull, Thomas~N Kipf, Peter Bloem, Rianne Van Den~Berg, Ivan
  Titov, and Max Welling.
\newblock Modeling relational data with graph convolutional networks.
\newblock pages 593--607, 2018.

\bibitem{schuld2020circuit}
Maria Schuld, Alex Bocharov, Krysta~M Svore, and Nathan Wiebe.
\newblock Circuit-centric quantum classifiers.
\newblock {\em Physical Review A}, 101(3):032308, 2020.

\bibitem{schuld2018supervised}
Maria Schuld and Francesco Petruccione.
\newblock Supervised learning with quantum computers.
\newblock {\em Quantum science and technology}, 17, 2018.

\bibitem{schuld2021machine}
Maria Schuld and Francesco Petruccione.
\newblock {\em Machine learning with quantum computers}, volume 676.
\newblock Springer, 2021.

\bibitem{tuysuz2021hybrid}
Cenk T{\"u}ys{\"u}z, Carla Rieger, Kristiane Novotny, Bilge Demirk{\"o}z,
  Daniel Dobos, Karolos Potamianos, Sofia Vallecorsa, Jean-Roch Vlimant, and
  Richard Forster.
\newblock Hybrid quantum-classical graph neural networks for particle track
  reconstruction.
\newblock {\em Quantum Machine Intelligence}, 3(2):29, 2021.

\bibitem{velickovic2019deep}
Petar Veli{\v{c}}kovi{\'c}, William Fedus, William~L Hamilton, Pietro Li{\`o},
  Yoshua Bengio, and R~Devon Hjelm.
\newblock Deep graph infomax.
\newblock {\em arXiv preprint arXiv:1809.10341}, 2018.

\bibitem{verdon2019quantum}
Guillaume Verdon, Trevor McCourt, Enxhell Luzhnica, Vikash Singh, Stefan
  Leichenauer, and Jack Hidary.
\newblock Quantum graph neural networks.
\newblock {\em arXiv preprint arXiv:1909.12264}, 2019.

\bibitem{xu2018powerful}
Keyulu Xu, Weihua Hu, Jure Leskovec, and Stefanie Jegelka.
\newblock How powerful are graph neural networks?
\newblock {\em arXiv preprint arXiv:1810.00826}, 2018.

\bibitem{yang2016revisiting}
Zhilin Yang, William Cohen, and Ruslan Salakhudinov.
\newblock Revisiting semi-supervised learning with graph embeddings.
\newblock In {\em International conference on machine learning}, pages 40--48.
  PMLR, 2016.

\bibitem{ye2025quantum}
Zi~Ye, Kai Yu, and Song Lin.
\newblock Quantum graph convolutional networks based on spectral methods.
\newblock {\em arXiv preprint arXiv:2503.06447}, 2025.

\bibitem{zheng2021quantum}
Jin Zheng, Qing Gao, and Yanxuan L{\"u}.
\newblock Quantum graph convolutional neural networks.
\newblock In {\em 2021 40th Chinese Control Conference (CCC)}, pages
  6335--6340. IEEE, 2021.

\bibitem{zheng2024quantum}
Jin Zheng, Qing Gao, Maciej Ogorza{\l}ek, Jinhu L{\"u}, and Yue Deng.
\newblock A quantum spatial graph convolutional neural network model on quantum
  circuits.
\newblock {\em IEEE Transactions on Neural Networks and Learning Systems},
  36(3):5706--5720, 2024.

\end{thebibliography}

\end{document}